\documentclass[runningheads]{llncs}

\usepackage[T1]{fontenc}
\usepackage{cite}
\usepackage{amsmath,amssymb,amsfonts}
\usepackage{graphicx}
\usepackage{booktabs}
\usepackage{setspace}
\usepackage{tikz}
\usepackage{listings}
\usepackage{microtype}
\usepackage{hyperref}
\usetikzlibrary{arrows.meta,positioning,shadows,fit,calc}

\title{Promise-Future Synchronization for Cluster Asynchronous Many-Task Runtimes via\\MPI One-Sided Communication}

\author{Mia Reitz}

\authorrunning{M. Reitz}
\titlerunning{Promise-Future Synchronization for Cluster AMT}

\institute{University of Kassel, \textit{Research Group Programming Languages/Methodologies}\\
\email{mia.reitz@uni-kassel.de}
}

\begin{document}
\maketitle

\begin{abstract}
Asynchronous Many-Task~(AMT) runtimes use futures as placeholders for values produced by other tasks.
In the ItoyoriFBC AMT runtime, the existing \emph{future-only} model binds each future to its producer at creation time and requires the number of tasks that read each future to be fixed at compile time.
This prevents directly expressing algorithms that create dependencies dynamically.
We extend ItoyoriFBC with an implementation of a \emph{promise-future} model that lifts these limitations.
Thereby, our ItoyoriFBC variant supports dynamic algorithms such as Hierarchical LU factorization~(HLU).
We experimentally evaluated our implementation using HLU on up to 16~nodes and observed near-ideal scaling with a $15.6\times$~speedup.
\keywords{Futures \and Promises \and Work Stealing \and Asynchronous Many-Task Programming}
\end{abstract}

\section{Introduction}
Asynchronous Many-Task~(AMT) runtimes (e.g., HPX and Legion) divide computations into fine-grained tasks and often express dependencies through \emph{futures}: placeholders for values that producer tasks create and consumer tasks read later.
A runtime typically assigns the tasks dynamically to worker processes through work stealing, where idle workers steal tasks from others.

Itoyori~\cite{itoyori} is a cluster AMT runtime combining work stealing with MPI one-sided communication for nested fork-join programs, and ItoyoriFBC extends it with support for futures~\cite{MiaFutureSideEffects}.
The futures in ItoyoriFBC follow a \emph{future-only} model that binds each future to its producer task during task creation and requires a compile-time fixed consumer count (see Listing~\ref{lst:old}).
The future-only model works when dependencies are known at task creation, but not when tasks discover dependencies dynamically while building the task graph such as in a recent algorithm for the LU factorization of hierarchical matrices~(HLU)~\cite{ruedigerHLU}.

The \emph{promise-future} model separates the producer side from the consumer side: a \emph{promise} provides the value, while \emph{futures} consume it (Listing~\ref{lst:new}).
For example, the recent HLU algorithm requires a parent task to pass futures to child tasks before spawning the corresponding producer tasks.

Realizing this model in ItoyoriFBC creates two challenges.
First, the runtime must know when it is safe to reclaim the memory of task results, even though the number of futures may vary during execution.
Second, a consumer task may access a future before the corresponding producer task exists, so the runtime must be able to pause and later resume a variable number of consumers (called \emph{waiters}).
This work investigates how the promise-future model can be realized in ItoyoriFBC.
Our contributions are: (1)~a data structure design to handle the two challenges, (2)~its implementation in ItoyoriFBC, and (3)~an evaluation of our implementation with HLU on up to 16~nodes.%
\begin{figure}[!htbp]
\begin{minipage}[t]{0.44\textwidth}
\begin{lstlisting}[caption={Future-only Model\label{lst:old}}]
future<int, 1> f = 
  spawn([=] { return 42; });
int v = f.get();
\end{lstlisting}
\end{minipage}\hfill
\begin{minipage}[t]{0.5\textwidth}
\begin{lstlisting}[caption={Promise-Future Model\label{lst:new}}]
promise<int> p = promise<int>::make();
future<int> f = p.get_future();
spawn([p] { p.set(42); });
int v = f.get();
\end{lstlisting}
\end{minipage}
\end{figure}%
\section{Design}
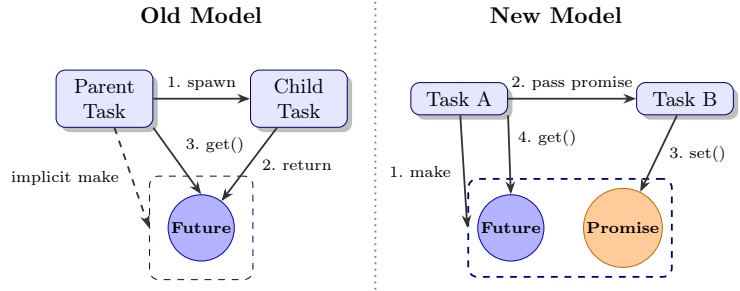
\begin{figure}[!tbp]
  \centering
  \resizebox{0.8\linewidth}{!}{\begin{tikzpicture}[
  font=\footnotesize,
  task/.style={draw, fill=blue!10, drop shadow, rounded corners, align=center, inner sep=4pt, minimum width=1.5cm, draw=blue!50!black},
  handle/.style={draw, circle, inner sep=1pt, minimum size=0.6cm, font=\scriptsize\bfseries},
  arrow/.style={-{Stealth[length=2mm]}, line width=0.8pt, draw=black!80},
  label/.style={font=\scriptsize\itshape, text=black}
]

  \node[font=\bfseries] (label_old) at (1.5, 3.8) {Old Model};
  
  \node[task] (parent) at (0, 2.5) {Parent\\Task};
  \node[task, right=1.5cm of parent] (child) {Child\\Task};
  
  \node[handle, fill=blue!30, draw=blue!80!black] (fut) at (1.5, 0.5) {Future};

  \node[draw=blue!50!black, dashed, rounded corners, fit=(fut), inner sep=8pt, label={[font=\tiny, text=blue!80!black]below:Shared State}] (state_old) {};
  
  \draw[arrow] (parent) -- node[above, font=\scriptsize] {1. spawn} (child);
  
  \draw[arrow, dashed] (parent) -- node[left=2pt, font=\scriptsize] {implicit make} (state_old.west);
  
  \draw[arrow] (child) -- node[right=2pt, font=\scriptsize] {2. return} (fut);
  
  \draw[arrow] (parent.south east) -- node[above right, font=\scriptsize] {3. get()} (fut.north);

  \draw[dotted, thick, gray] (4.2, 4.0) -- (4.2, -0.5);

  \begin{scope}[xshift=5.5cm]
    \node[font=\bfseries] (label_new) at (1.5, 3.8) {New Model};

    \node[task] (taskA) at (0, 2.5) {Task A};
    \node[task] (taskB) at (3.5, 2.5) {Task B};
    
    \node[handle, fill=blue!30, draw=blue!80!black] (fut_new) at (0.8, 0.5) {Future};
    \node[handle, fill=orange!40, draw=orange!80!black, right=0.6cm of fut_new] (prom) {Promise};
    
    \node[draw=blue!50!black, dashed, thick, rounded corners, fit=(prom)(fut_new), inner sep=4pt, label={[font=\tiny, text=blue!80!black]below:Shared State P}] (state) {};

    \draw[arrow] (taskA) -- node[above, font=\scriptsize] {2. pass promise} (taskB);
    
    \draw[arrow] (taskA) -- node[left=2pt, font=\scriptsize] {1. make} (state.west);
    
    \draw[arrow] (taskB) -- node[right=1pt, font=\scriptsize] {3. set()} (prom);
    
    \draw[arrow] (taskA.south east) -- node[above right, font=\scriptsize] {4. get()} (fut_new.north);
  \end{scope}

\end{tikzpicture}}
  \caption{Task interaction sequence: future-only vs.\ promise-future model.}
  \label{fig:interaction}
\end{figure}

As shown in Figure~\ref{fig:state}, futures in ItoyoriFBC use a \emph{shared state} object that stores the produced value and tracks tasks that may access or wait for it.
In our extension, each promise and its corresponding futures refer to this shared state~(see Figure~\ref{fig:interaction}), stored on a \emph{home process}, i.e., the MPI process responsible for it. The home process creates the promise; ideally, it also accesses the future data, since each access may require MPI one-sided communication.

\begin{figure}[!tbp]
  \begin{minipage}[b]{0.46\textwidth}
    \centering
    \resizebox{\linewidth}{!}{\begin{tikzpicture}[
  font=\normalsize,
  box/.style={draw, rounded corners, align=center, inner sep=5pt, minimum width=4.1cm, minimum height=1.5cm, line width=0.8pt},
  handle/.style={draw, rounded corners, align=center, inner sep=3pt, minimum width=1.8cm, font=\bfseries},
  oldbox/.style={box, fill=blue!10, draw=blue!60!black},
  newbox/.style={box, fill=blue!10, draw=blue!60!black},
  future/.style={handle, fill=blue!20, draw=blue!70!black},
  promise/.style={handle, fill=orange!20, draw=orange!60!black},
  arrow/.style={-{Stealth[length=2.0mm]}, line width=0.8pt, draw=black!80}
]
  \node[oldbox] (old) {Shared State};
  \node[newbox, right=1.1cm of old] (new) {Shared State};

  \node[future, above=8mm of old, xshift=-1.1cm] (oldf1) {Future};
  \node[future, above=8mm of old, xshift=1.1cm] (oldf2) {Future};
  \draw[arrow] (oldf1) -- (old);
  \draw[arrow] (oldf2) -- (old);

  \node[future, above=8mm of new, xshift=-1.1cm] (newf1) {Future};
  \node[future, above=8mm of new, xshift=1.1cm] (newf2) {Future};
  \node[promise, below=8mm of new] (newp) {Promise};
  \draw[arrow] (newf1) -- (new);
  \draw[arrow] (newf2) -- (new);
  \draw[arrow] (newp) -- (new);
\end{tikzpicture}}
  \end{minipage}\hfill
  \begin{minipage}[b]{0.46\textwidth}
    \centering
    \resizebox{\linewidth}{!}{\begin{tikzpicture}[
  font=\small,
  event/.style={draw, rounded corners, align=center, inner sep=4pt, minimum width=2.2cm, minimum height=1.0cm, fill=blue!10, draw=blue!50!black, line width=0.8pt},
  promise_event/.style={event, fill=orange!10, draw=orange!50!black},
  arrow/.style={-{Stealth[length=2.0mm]}, line width=0.8pt, draw=black!80},
  label/.style={font=\footnotesize\bfseries, text=black},
]
  \node[promise_event] (e1) {Create promise\\count=1};
  \node[event, right=0.4cm of e1] (e2) {Create future\\count=2};
  \node[promise_event, right=0.4cm of e2] (e3) {Set value\\wake waiters};

  \node[promise_event, below=0.5cm of e3] (e4) {Destroy promise\\count=1};
  \node[event, left=0.4cm of e4] (e5) {Destroy future\\count=0 (GC)};

  \draw[arrow] (e1) -- (e2);
  \draw[arrow] (e2) -- (e3);
  \draw[arrow] (e3) -- (e4);
  \draw[arrow] (e4) -- (e5);

  \node[label, above=0.1cm of e1] {Start};
  \node[label, below=0.1cm of e5] {End};
\end{tikzpicture}}
  \end{minipage}

  \begin{minipage}[t]{0.46\textwidth}
    \caption{Shared state object: future-only (left) vs.\ promise-future model (right).}
    \label{fig:state}
  \end{minipage}\hfill
  \begin{minipage}[t]{0.46\textwidth}
    \caption{Reference counting of the promise-future model.}
    \label{fig:refcount}
  \end{minipage}
\end{figure}

\paragraph{Challenge 1: Reclaiming Shared State}
Because futures may be created dynamically (e.g., by copying a future), the runtime can no longer statically know when a shared state becomes unreachable.
We therefore attach a reference counter to each shared state: creating a future or promise increments it, and destroying one decrements it.
When the counter reaches zero, no task can access the shared state, so the home process may reclaim it (Figure~\ref{fig:refcount}): instantly if the home process performed the final decrement, otherwise at its next shared-state allocation.

\paragraph{Challenge 2: Suspending and Resuming Consumers}
If a consumer task accesses a future before its value is ready, or before the corresponding producer task exists, the runtime must suspend the consumer and later resume all waiters.
The runtime stores the consumer's continuation, i.e., the information needed to resume the consumer later, in a waiter list attached to the shared state.
Because the producer may run on a different process than the home process, the waiter list is a lock-free linked list accessed via MPI one-sided communication without involvement of the home process.
Listings~\ref{lst:push} and~\ref{lst:wake_all} provide pseudocode of the above two waiter list procedures:

\texttt{push:} When a waiter suspends itself, the waiter allocates a list node~(Line~2) and uses remote atomic Compare-And-Swap~(CAS) to push it onto the waiter list~(Lines~6--7); if the producer has already fulfilled the value, the CAS fails and the consumer resumes immediately~(Line~8).

\texttt{wake\_all:} When a producer writes a value to its shared state, the producer uses CAS to detach the waiter list, replacing the head with \texttt{MARKED\_DONE}~(Lines~4--5), and then traverses the detached list to wake suspended tasks~(Lines~9--12).

\begin{figure}[!tbp]
\begin{minipage}[t]{0.47\textwidth}
\begin{lstlisting}[caption={Pseudocode of \texttt{push}\label{lst:push}}]
bool push(State* S, Continuation c) {
  Node* n = new Node(c);
  Node* head = read_remote(S->head);
  while (head != MARKED_DONE) {
    n->next = head;
    Node* old = CAS_remote(
      &S->head, head, n);
    if (old == head) return true;
    head = old;
  }
  delete n;
  return false;
}
\end{lstlisting}
\end{minipage}\hfill
\begin{minipage}[t]{0.455\textwidth}
\begin{lstlisting}[caption={Pseudocode of \texttt{wake\_all}\label{lst:wake_all}}]
void wake_all(State* S) {
  Node* head = read_remote(S->head);
  while (true) {
    Node* old = CAS_remote(
      &S->head, head, MARKED_DONE);
    if (old == head) break;
    head = old;
  }
  while (head) {
    wake(head);
    head = head->next;
  }
}
\end{lstlisting}
\end{minipage}
\end{figure}

\section{Experiments}
\begin{figure}[!tbp]
  \centering
  \includegraphics[width=0.48\linewidth]{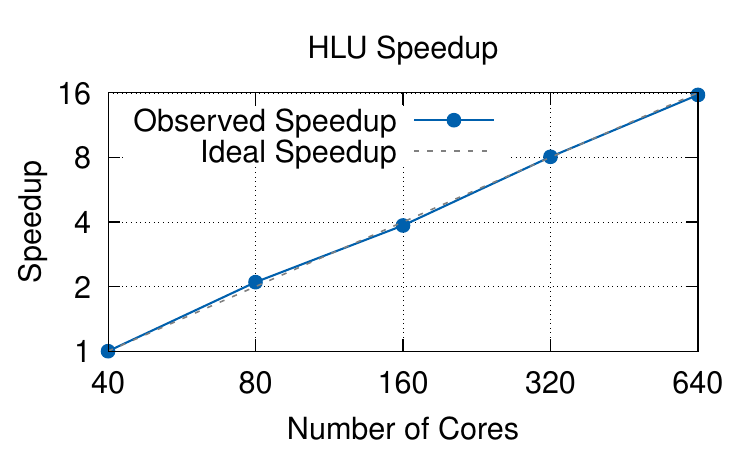}
  \caption{HLU speedup relative to one node on up to 16 nodes.}
  \label{fig:speedup}
\end{figure}

We evaluated our implementation with the HLU benchmark from~\cite{wamta25Ruediger}, using a full quadtree of height~$7$ (21\,845 tasks) and calibrated busy-waits to model 100 seconds of numerical work. Smaller instances ran too briefly to be meaningful, and larger ones exceeded cluster memory.
Experiments ran on the Goethe cluster~\cite{ClusterGoethe} of University of Frankfurt with Open MPI~5.0.5 and g++~11.4.1~(\texttt{-O3}), using 1--16~nodes (40--640~cores) and 10 runs per configuration.
We ran one worker per core.
Here, speedup is defined as the one-node running time divided by the running time on $n$ nodes.
Figure~\ref{fig:speedup} shows that HLU reaches $15.6\times$~speedup on 16~nodes with our variant; the variance across the 10 runs was low.

\section{Conclusions}
We realized the promise-future model in ItoyoriFBC using distributed reference counting and an MPI one-sided waiter list, and demonstrated that HLU reaches $15.6\times$~speedup on 16~nodes with our variant.
Future work should improve the reference counting to aggregate increments and decrements and only inform the home process when needed, and compare against the original future-only implementation.

\medskip

\noindent
\textbf{Acknowledgements:} 
This research was funded by the Deutsche Forschungsgemeinschaft (DFG, German
Research Foundation) under project number 512078735.
The author gratefully acknowledges the computing time provided to them on the Goethe-NHR cluster at the Frankfurt Center for Scientific Computing.
We also thank John Hundhausen for his assistance with code refactoring.

\bibliographystyle{splncs03_unsrt}
\bibliography{bibo}
\end{document}